\documentclass[sigconf, authorversion=true, nonacm=true]{acmart}

\AtBeginDocument{%
  \providecommand\BibTeX{{%
    \normalfont B\kern-0.5em{\scshape i\kern-0.25em b}\kern-0.8em\TeX}}}

\copyrightyear{2021}
\acmYear{2021}
\setcopyright{acmlicensed}\acmConference[WebSci '21]{13th ACM Web Science
Conference 2021}{June 21--25, 2021}{Virtual Event, United Kingdom}
\acmBooktitle{13th ACM Web Science Conference 2021 (WebSci '21), June
21--25, 2021, Virtual Event, United Kingdom}
\acmPrice{15.00}
\acmDOI{10.1145/3447535.3462549}
\acmISBN{978-1-4503-8330-1/21/06}

\usepackage{balance}
\usepackage{algorithm}
\usepackage[noend]{algpseudocode}

\newcommand{\textt}[1]{\texttt{#1}}

\begin{document}

\title{Wide-AdGraph: Detecting Ad Trackers with a Wide Dependency
Chain Graph}


\author{Amir Hossein Kargaran}
\authornote{Department of Electrical and Computer Engineering, Isfahan University of Technology, Isfahan 8415683111, Iran.}
\affiliation{%
  \institution{Isfahan University of Technology}
  \city{}
  \country{}
}
\email{kargaran@ec.iut.ac.ir}

\author{Mohammad Sadegh Akhondzadeh}
\authornotemark[1]
\affiliation{%
  \institution{Isfahan University of Technology}
  \city{}
  \country{}}
\email{ms.akhondzadeh@ec.iut.ac.ir}

\author{Mohammad Reza Heidarpour}
\authornotemark[1]
\affiliation{%
  \institution{Isfahan University of Technology}
  \city{}
  \country{}}
\email{mrheidar@iut.ac.ir}

\author{Mohammad Hossein Manshaei}
\authornotemark[1]
\affiliation{%
  \institution{Isfahan University of Technology}
  \city{}
  \country{}}
\email{manshaei@iut.ac.ir}

\author{Kave Salamatian}
\authornote{LISTIC – Polytech Annecy-Chambéry, BP 80439, 74944 Annecy le Vieux Cedex, France.}
\affiliation{%
  \institution{University of Savoie Mont Blanc}
  \city{}
  \country{}}
\email{kave.salamatian@univ-smb.fr}

\author{Masoud Nejad Sattary}
\authornote{Chavoosh ICT, No. 4, Ferdowsi St., Isfahan, Iran.}
\affiliation{%
  \institution{Chavoosh ICT}
  \city{}
  \country{}}
\email{sattary@chavoosh.com}

\renewcommand{\shortauthors}{Kargaran et al.}
\renewcommand{\shorttitle}{Wide-AdGraph}

\begin{abstract}
Websites use third-party ads and tracking services to deliver targeted ads and collect information about users that visit them. These services put users' privacy at risk, and that is why users' demand for blocking these services is growing.
Most of the blocking solutions rely on crowd-sourced filter lists manually maintained by a large community of users. In this work, we seek to simplify the update of these filter lists by combining different websites through a large-scale graph connecting all resource requests made over a large set of sites.
The features of this graph are extracted and used to train a machine learning algorithm with the aim of detecting ads and tracking resources.
As our approach combines different information sources, it is more robust toward evasion techniques that use obfuscation or changing the usage patterns. We evaluate our work over the Alexa top-10K websites and find its accuracy to be 96.1\% biased and 90.9\% unbiased with high precision and recall. It can also block new ads and tracking services, which would necessitate being blocked by further crowd-sourced existing filter lists. Moreover, the approach followed in this paper sheds light on the ecosystem of third-party tracking and advertising.
\end{abstract}


\maketitle

\section{Introduction}
The underlying business model of Internet growth, which is mainly based on free access to online resources, has driven a full industry of targeted advertisements that generate revenue for the content providers. These advertisements and related services are delivered through online services, that provide customized ads or web analytics~\cite{lerner2016internet} by often inserting \emph{scripts} and sub-documents, {\em i.e.}, \emph{Iframes,} in web pages. These advertisement services are generally backed by tracking services that harvest users' behavior during their browsing experience in order to gather information for generating user profiles that will be leveraged to provide customized and targeted ads to the user. In the forthcoming,
we will use the term \emph{first-party} to refer to the targeted website the user is visiting and use the term "{\em AdTracker}" to refer to both advertisement and added value services that are backed by trackers. We will use the term \emph{third-parties} to refer to any domain contacted by a first-party website that is different from its source domain. Some of these third-parties are among AdTrackers, and some of them are not. The ultimate goal is to classify the AdTrackers among third-parties. 

While the advertisement industry and its related personal data harvesting through tracking have become a fundamental component of nowadays Internet, they come at a considerable cost of disregarding users' privacy. They are also considered by the users as a nuisance, increasing the bandwidth and degrading the browsing experience considerably. 
In recent years, many researchers engaged in analyzing the ads and tracking ecosystem and understanding the means and techniques used by AdTrackers to gather personal data. They provide targeted advertisements with the ultimate goal of preserving user's privacy \cite{cozza2020hybrid}. Several ad-blocking tools, such as \textit{Ghostery}~\cite{Ghostery}, \textit{No Script}~\cite{noscript}, \textit{Adblock}~\cite{Adblock} and \textit{Adblock Plus}~\cite{Adblockplus}, \textit{Disconnect}~\cite{Disconnect}, \textit{Privacy Badger}~\cite{Pbadger}, {\em etc.}, have been developed. These tools generally leverage filter lists that are frequently created through a crowd-sourcing of a users' community~\cite{wills2016ad}. AdTrackers are detected through matching rules defined over features that can be observed in a browser. AdTrackers frequently change their behavior to evade detection and continuously develop new tracking techniques~\cite{adwars}, and the detection filter lists and rulesets have to be regularly updated to remain effective~\cite{vastel2018filters}. Moreover, coverage of popular AdBlock lists might not be uniformly complete as AdTrackers in some geographic and linguistic areas might be less frequently reported \cite{falahrastegar2014anatomy, cartography,hashmi2019longitudinal,sjosten2020filter}.
All these elements point toward the importance of developing automated methods for detecting both the new behaviors of already known AdTrackers and the emergence of unknown actors. This entails efficiently managing the filter ruleset by inserting novel rules or eventually removing the obsolete ones. 

Automatic detection of AdTrackers using their observed activities in relation to a site is a challenging task.
There are three main approaches to this problem. In the first approach, we can use some specific features to give useful indications, {\em e.g.}, the observation rate among different websites, {\em i.e.}, third-party AdTrackers are generally observed in several websites. Indeed, the problem becomes to choose appropriate thresholds for the observation rates of different features to decide whether an AdTracker exists or not. In the second approach, the specific behaviors of trackers are leveraged. In particular, trackers have to send the collected information to their servers, generating generic traces that can be used to identify them. The third and more powerful approach exploits the web of relationships among third-parties. Advertisers and trackers have to interact in order to exchange information.
Moreover, trackers also exchange information in order to grow their business and access more data. Uncovering this interaction network helps in detecting AdTrackers, and it can provide a glance at the global ecosystem of AdTrackers. However, building this web of relationships entails large-scale data collection and analysis.

This paper mainly targets the third approach described above. We will define a single large {\em web graph} capturing dependencies between AdTrackers. This graph is built through web crawling and adding newly observed AdTrackers. We will also show how this graph can provide insights into the ecosystem of AdTrackers and how it can be used to detect and label new AdTrackers by a supervised machine learning model. The  contributions in this paper can be summarized as follows:
\begin{enumerate}
    \item We define a directed graph, named \textbf{wide AdTrackers' graph (\emph{Wide-AdGraph})} which is a dependency chain that models the relationship between AdTracker services. We present a measurement method to build this graph and use it to extract some insights into the AdTracker ecosystem.
    \item We develop a {\bf supervised representation learning algorithm} that combines structural properties extracted from the {\em Wide-AdGraph}, with content features in order to classify AdTrackers. As a side to this, we have developed an \textbf{effective keywords extraction method} that retrieves the most useful keywords from URL requests.
    \item We show that after training, the classifier can retrieve AdTrackers which are also detected by the existing filter list,  with a 96.1\% biased and 90.9\% unbiased accuracy. Moreover, we uncover some unreported new AdTrackers. 
\end{enumerate}
The rest of the paper is organized as follows.
Section~\ref{sec-back}, introduces the background and describes the dataset collection. Section~\ref{sec-graphmaking}, describes {\em Wide-AdGraph} model and its features. Section~\ref{sec-features} shows featurization with two types of features from {\em Wide-AdGraph} for the detection of AdTrackers, in detail.
Section~\ref{sec-classification} presents the AdTracker classification model that is evaluated in Section~\ref{sec-evaluation}.
A comparison with the state-of-the-art in the field is described in Section~\ref{sec-related}. Finally, we give our final remarks in Section~\ref{sec-conclusion}.

\section{Background and Dataset}\label{sec-back}
A user connecting to a website has to provide a lot of information that is needed for the operation of the website, like IP address, browser parameters, {\em etc.} However, this information might be considered private, as it enables
fingerprinting and tracing the users \cite{6547132}. This generates a privacy problem when the website (first-party) leaks some of the user connection information to third-parties. This leakage can happen consciously to benefit from advertisement revenues or unconsciously by embedding services that seem to be beneficial to the user or the content provider, {\em e.g.}, web analytics, but enables a third-party to gather personal information. Eventually, the "data brokers" can harvest fine grains and precise information about the users by combining information obtained over a large number of websites. This enables them to profile individual user and harm privacy ~\cite{anthes2014data,lerner2016internet,englehardt2016online} and security \cite{li2012knowing,gervais2017quantifying,merzdovnik2017block}, along with reducing users satisfaction~\cite{PageFair} and performance \cite{garimella2017ad,pujol2015annoyed}. 

\subsection{Web Tracking dependency graph}
\begin{figure}[t]
\centering
\includegraphics[scale=0.38]{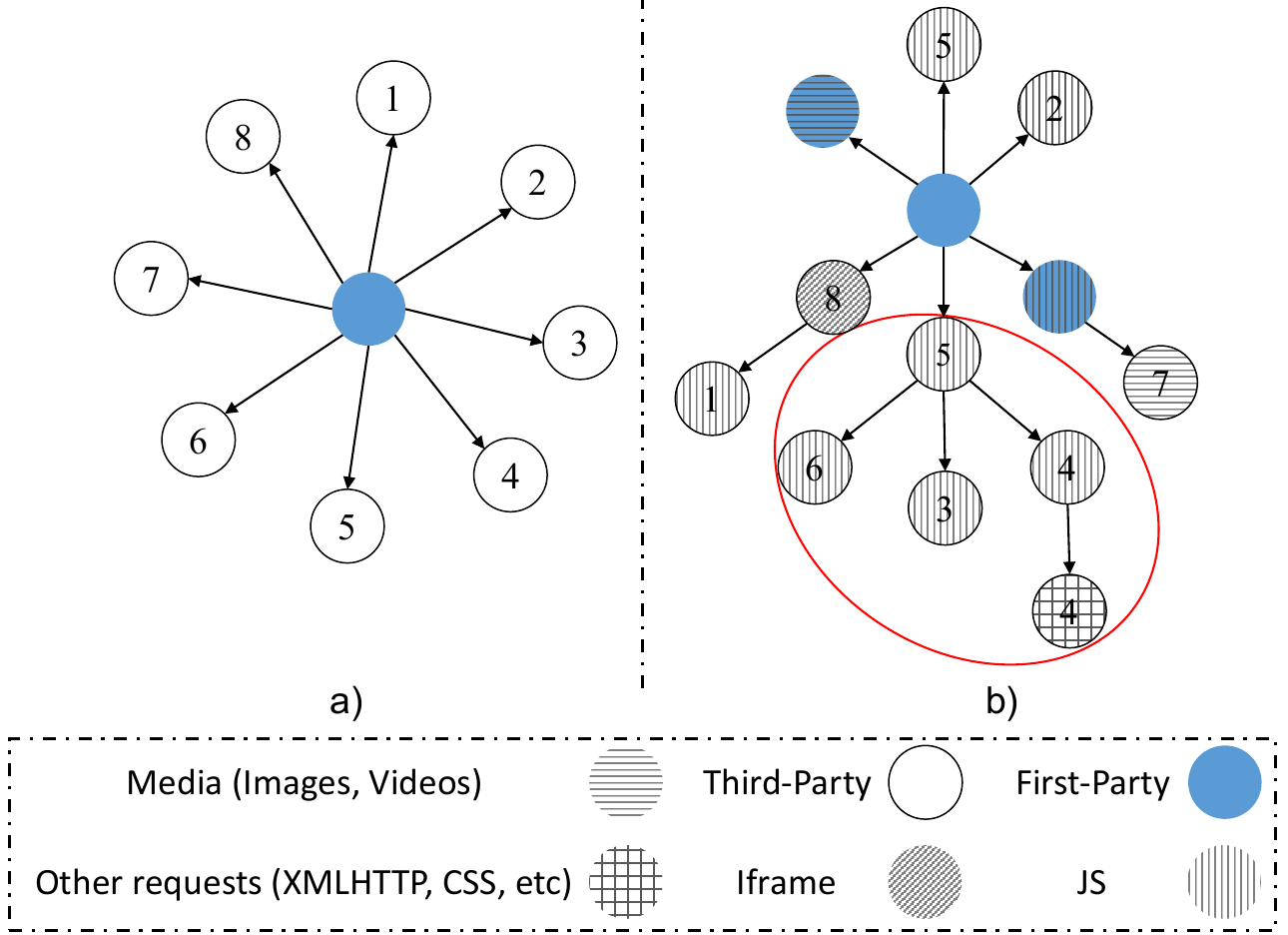}
\caption[a) Star bipartite graph VS b) Dependency chain graph]
{(a) Star bipartite graph vs (b) Dependency chain graph for a website: Each number represents a unique domain. In a bipartite graph, all third-party services are connected by nodes connected first-party root. However, in the dependency chain, nodes that depend on other nodes are linked via a directed edge.}
\label{fig-digichain}
\end{figure}

Trackers can harvest information through different means. A popular method is to use Javascripts and Iframes because of the ease of their placement in first-party codes. A script added by a third-party might contact its home server or connect to several other third-parties and send them information. It might even load additional JavaScripts to extract information. These bounced data transfers make third-party detection more challenging as they are not directly seen in the first-party code. For example, in Fig.~\ref{fig-digichain}-a we show a bipartite graph without considering bounced third-party relations for a website.
Nevertheless, as JavaScript and Iframe runs in the client browser, they are detectable. The calling relationship among downloaded objects in a web session creates a dependency chain \cite{ikram2019chain} that can be represented through a directed graph. For example, in Fig.~\ref{fig-digichain}-b, we show a dependency graph for a website containing a JavaScript from domain 5 that loads three other scripts from domains 3, 4, 6, and the JavaScript from domain 4  making a request to domain 4 servers. Throughout this paper, we will refer by "domain" to the second-level of DNS domain name, and by sub-domain to the third-level of DNS name \footnote{We have used the \textit{tldextract} tool to extract the domain levels of URLs~\cite{tldextract}}.

These dependencies chain have already been exploited in some way in prior works \cite{ikram2017towards, wu2016machine,sjosten2020filter,iqbal2020adgraph,Disconnect,lightbeam}. In \cite{Disconnect,lightbeam} JavaScript and HTTP requests are exploited to build a star graph for each website. The nodes in this graph are the source domain of executed JavaScripts along with domains that are requested through HTTP requests and might have further labels relative to their role (as AdTrackers). Links represent the fact that this domain has been called during the web session. The Links might also be labeled by the type of calls made to the corresponding domains, {\em e.g.}, script download, HTTP requests, CSS requests, {\em etc.}  An example of such a graph is shown in Fig.~\ref{fig-digichain}-a. However, such graph can be refined. One can enrich the graph by maintaining the precise calling dependency, {\em i.e.}, not directly attaching a domain called after a bounced call to the website, as is done in Fig.~\ref{fig-digichain}-b. This results in a tree rather than a star that is rooted in each website. The tree-based view provides a more detailed observation of the dependencies that will be proved to be beneficial for AdTracker classification. Such graphs have been used in \cite{iqbal2020adgraph,sjosten2020filter} and are maintained separately for each website. In \cite{lightbeam}, star-like graphs are merged together, while in this paper, we will merge trees coming from all websites into a large-scale global dependency graph as we will describe in Sec. \ref{sec-graphmaking}.

\subsection{Dataset}
In order to extract dependencies and construct a graph, we have used an instrumented \emph{Chromium} browser controlled with the \emph{Selenium}~\cite{Selenium} browser automation tool. Through these tools, we can monitor all scripts executed locally and detect the execution of bounced scripts. The \emph{Chromium}'s developer tool has a useful feature that makes it possible to store all HTTP interactions between the client and destination website in the \emph{har} archival format ~\cite{devtool}. In this format, each requested object in a web session has an "initiator" element that describes the web resource that has initiated the requested object. This element defines the dependencies we are using to build our graph. Moreover, Chromium's web driver can adapt the downloading timeouts set to ensure that each web page is rendered completely (by default, 200 seconds). We checked the rendering status for all the websites and increased the waiting time for stalled crawls accordingly. We also used scrolling on web pages in order to counter bot-mitigation strategies implemented by sites.

Since we need proper ground truth and in order to compare with other papers, for evaluation purpose, we crawled, as it is done in previous works \cite{iqbal2020adgraph, sjosten2020filter}: the homepage of the web domains in the list of 10K websites with the highest number of visit \cite{alexatop}, {\em i.e.}, the top 10K Alexa lists. The instrumented browser was used to collect all web browser's interaction, {\em i.e.}, URLs, scripts, objects fetched by the browser, {\em etc.}, for all web pages. During the crawling, we observed tracker objects both embedded directly into web page code and generated dynamically by other objects. Indeed, the 10K Alexa web pages are not fully representative of the whole Internet ecosystem. However, it is noteworthy that each of the 10K websites only appears once in our dataset, differently from datasets obtained by looking at real people browsing history, resulting in heavy biases coming from the hugely skewed web sites' popularity distribution. We observed that around 10\% of websites in the list were not responsive and that some of these sites have implemented more aggressive bot-detection techniques that make them more difficult to analyze using Chromium and Selenium. Finally, in our final dataset, we attained a downloads success rate of 84\% of the top 10K Alexa websites, {\em i.e.}, our dataset finally contains 8,400 websites in the top 10K Alexa ranking.

\section{Wide-AdGraph Model}\label{sec-graphmaking}
In this section, we present \emph{Wide-AdGraph} and discuss its specific properties along with its suitability for discovering AdTrackers.

\begin{figure*}[t]
\centering
\includegraphics[width=\linewidth]{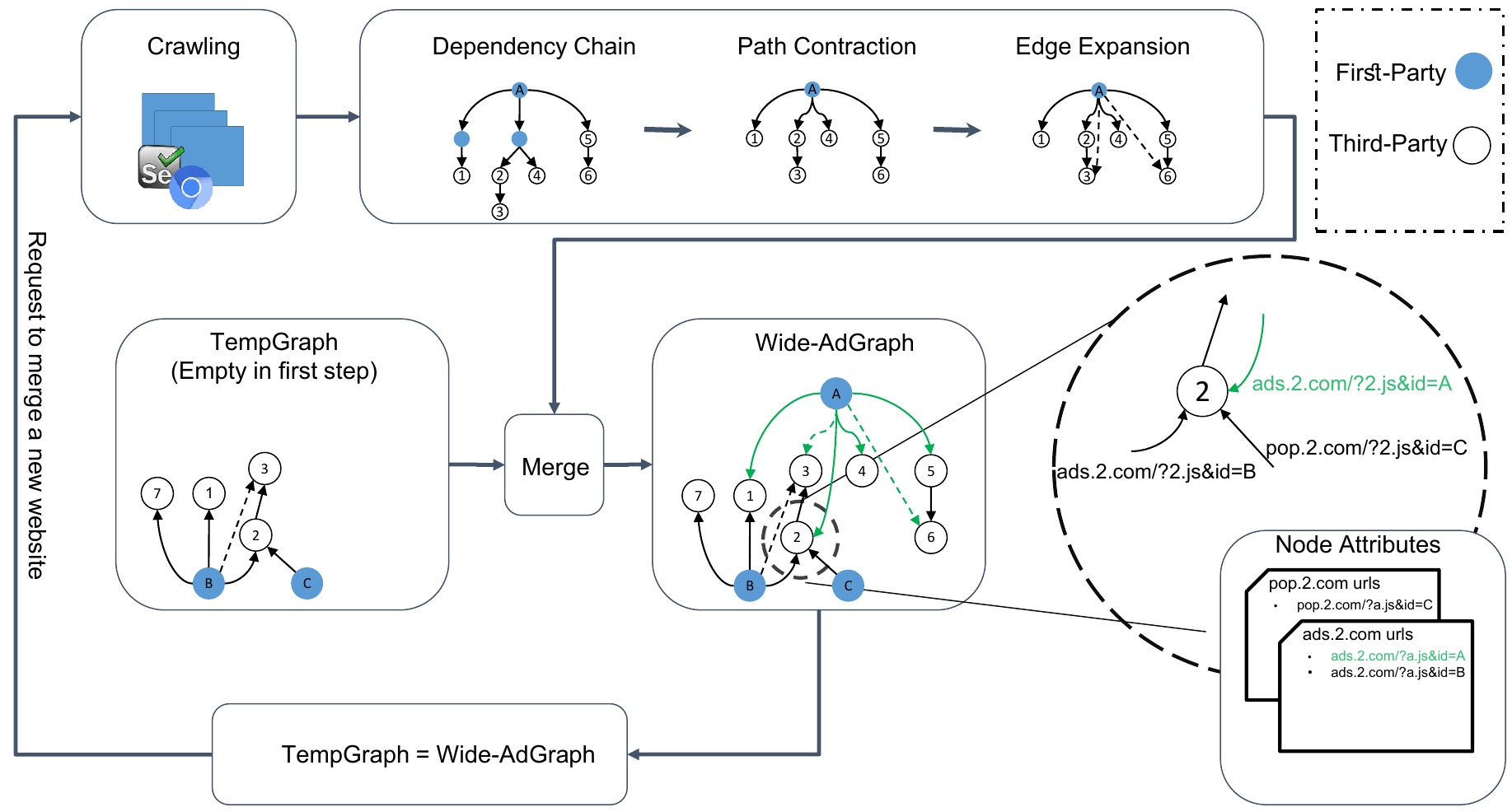}
\caption[Construction of {\em Wide-AdGraph}]
{\emph{Wide-AdGraph} construction. 
The procedure starts with crawling. The graph is expanded by requesting to crawl new websites. Each node with the same domain and interaction type is labeled with a unique number and the green edges indicate that they are new.}
\label{fig-graphmaking}
\end{figure*}

\subsection{Graph Construction}
As explained above, the har archive obtained for each visited website contains all URLs involved in the web session. They together generate several \textit{dependency chains} all rooted at the first-party URL. The question of how to choose the granularity of the graph's nodes is important. Some AdTrackers generate different sub-domains for each customer (as a kind of costumer's id), {\em e.g.}, marketo.com generates for each customer identified by \textit{\{munchkin-id\}} a specific URL in the form of \textit{\{munchkin-id\}.mktoresp.com} that will be used in order to upload tracking information. In such a case, the AdTracker node with the domain granularity has a low in-degree, while one might expect a high in-degree feature for AdTrackers. Moreover, a company might have some sub-domains that act as AdTrackers and some that are not, {\em e.g.}, \textit{adservice.google.com} is a AdTracker while \textit{translate.google.com} is not. However, we are not interested in dependencies inside a domain, as we can assume that data uploaded to a sub-domain will be shared with the whole domain. For this reason, and to reduce the size of the graph, we will consider each domain just a single node while maintaining a data structure, called a "{\em document}", containing relevant information for each sub-domain. This structure enables us to migrate between sub-domain and domain granularity when doing AdTracker classification. We have therefore regrouped all requests relative to each first-party domain into a single node. Therefore, each (super) node integrates all objects and requests relative to the labeled domain name. This regrouping results in a \textit{path contraction} as several nodes belonging to a dependency chain are regrouped into a single node if they all share the same domain. Each node in a dependency chain might  interact with the next node through four different forms :  \textt{Javascript}, \textt{Media} (Images, Videos),  \textt{Iframe}, or other types of requests, {\em e.g.}, XMLHTTP requests that are labeled as \textt{Other}, resulting into four different type of links that we called them as type. It is noteworthy that depending on the type of interaction present in a domain, only some of the above four types of interactions might be present in a node. Moreover, as described before, we  maintain the link not only to the node but also to the sub-domain documents. In fact, this is the combination of a document and a link pointing to it that gets classified into belonging to AdTracker or non-AdTracker categories.

In order to simplify the analysis and for reasons that will become clear later, we expand the graph described above by adding some virtual edges going from a visited web site to any node that is not directly connected to it, {\em i.e.}, bounced third-parties, and that belongs to one of its dependency chains. We label these virtual edges with \texttt{Bounced}.  We call this last step {\em edge expansion}. This results for each visited website into a directed graph around the visited website. We can merge these directed graphs obtained for all visited websites into a large graph. This is done by merging all nodes with the same domain into a single node with an edge set that is the union of all pre-merge edge sets, {\em i.e.}, two links with the same source and destination nodes and the same edge label are merged into a single one.  This directed graph is called the "{\em Wide-AdGraph}. We show all steps for building it in Fig.~\ref{fig-graphmaking} and Algorithm ~\ref{alg-graphmaking}.

\begin{algorithm}[t]
\caption{Construction of The Wide-AdGraph}
\label{alg-graphmaking}
\begin{algorithmic}[1]

\Procedure{MakeWide-AdGraph}{$Urls$} \\
     \Comment{Input: $List$ $Urls$}
     \Comment{Output:$Graph$ $Wide$\text{-}$AdGraph$}
\State $TempGraph(v,e) \gets Graph(\{\},\{\}) $
     
\For{each $url$ in $Urls$}
            \State {$ g(v,e) \gets \textit{MakeDepencencyChainGraph}(url)$}
            
             \State {$g(v,e) \gets g(v,e) / \{{x \rightarrow y | x,y \in g.first-party}\}$}
             \Comment {Path Contraction}
            
             \State {$g(v,e) \gets g(v,e \cup {\{{g.root \rightarrow \{v\} } \}})$ }
             \Comment{Edge Expansion}
            
             \State {$TempGraph \gets (g \cup Wide$\text{-}$AdGraph)$}
             \Comment{Merge}
             \State {$Wide$\text{-}$AdGraph \gets TempGraph$}         
\EndFor
       
\Return ($Wide$\text{-}$AdGraph$)

\EndProcedure
\end{algorithmic}
\end{algorithm}

\begin{figure*}[t]
\centering
\includegraphics[width=\linewidth]{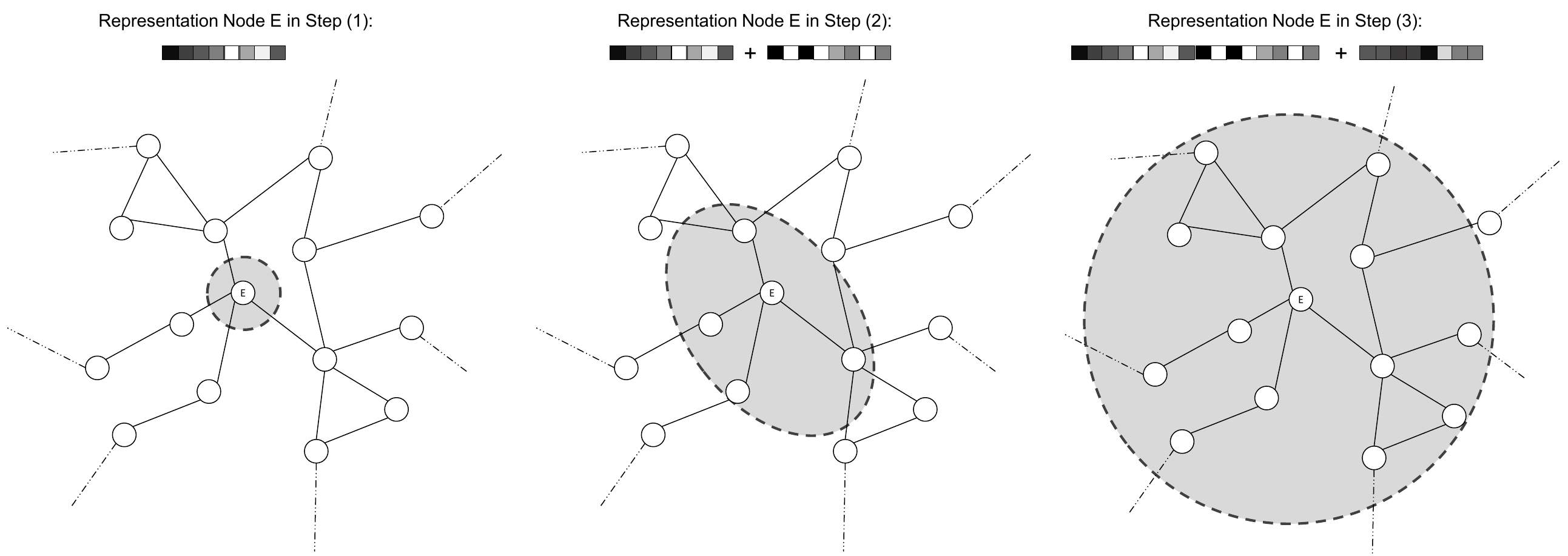}
\caption[Refex]
{Computing flow of representation for node E: Refex algorithm uses a recursive procedure to compute a representation for each node by aggregation of other node representations.}
\label{fig-refex}
\end{figure*}

\subsection{Wide-AdGraph {\em vs.} Other Graph Based Approaches}\label{subsec-adgraph}
There has been a significant set of prior works on automating AdTracker detection based on dependency graphs. We present some of the similarities and differences used in the dependency graphs. Adgraph~\cite{iqbal2020adgraph} and its extended work, Pagegraph~\cite{sjosten2020filter} use a graph that is generated similarly with what was described above for each source web site. However, Adgraph is not aggregating all these graphs into a single graph like Wide-AdGraph and uses each individual dependency graph separately. We will see later that using a single global graph reduces the bias strongly during the machine learning phase. Other attempts to build a dependency graph from AdTrackers have used the star-based representations (like in Fig~\ref{fig-digichain}-a) to form a bipartite graph, with first-parties on one side and third-parties on the other side  ~\cite{schelter2016ubiquity, kalavri2016like, lightbeam, Disconnect}. Some other works have built their graphs over the incomplete characterization of dependencies, {\em e.g.}, \cite{gomer2013network} only uses referral relations from domains to domains, and \cite{urban2018unwanted, bashir2018diffusion} uses direct first-party to third-party relations and third-parties cookie syncing relations. The {\em Wide-AdGraph} considers a complete representation of dependencies in the form of a tree with different node types (see Fig.~\ref{fig-digichain}-b) and merges local graphs into a large-scale one. Moreover, the {\em Wide-AdGraph} considers sub-domain granularity which has not been done before in previous papers. 

\subsection{Wide-AdGraph Properties}
The {\em Wide-AdGraph} combines all relevant information about the third-parties ecosystem that can be observed through a web browser into a single and holistic graph representation. Third-party nodes capture all the interactions and act as sinks for all harvested information. Moreover, each third-party domain and specific type of interaction appear at most once in the {\em Wide-AdGraph}. The out-degree of any first-party in the graph represents the number of AdTrackers that are contacted when a user gets it. Some of these third-parties are connected directly, while some are connected through other third-party intermediaries. The number of in-degree to an AdTracker node (only potential AdTrackers have incoming edges) represents how many other domains send information to it. We define for each third-party a "{\em direct coverage}" that represents the percentage of first-party nodes that are connected directly to it. We also define the "{\em indirect coverage}" value that represents the percentage of first-party nodes that have a path connecting them to this third-party. The direct coverage of an AdTracker shows the proportion of first-party sites where a user can be tracked through direct information transfer. The indirect coverage shows the potential portion of first-party sites where the AdTracker can track the user. Another metric of interest is also the average path length between each of the third-parties and first-parties.

\section{Feature Extraction}\label{sec-features}
We need to extract features from the {\em Wide-AdGraph} in order to use them as predictors in the classification. 
These features can be of two types: "Content features" that only depend on the individual interaction made between two adjacent nodes, and  "Structural features", which are obtained over the {\em Wide-AdGraph} and represents the global relationship between nodes in the graph. For example, in Fig.~\ref{fig-refex}, we show these relationships for multiple websites in {\em Wide-AdGraph}.
Both these attributes are gathered for each sub-domain and stored in relevant nodes of the Wide-AdGraph and in the related documents.


\subsection{\textbf{Content Features}}
Content features are obtained by looking at individual interactions between nodes in the {\em Wide-AdGraph}. The interactions between nodes consist of HTTP requests made to URLs that usually contain some specific keywords or patterns that help in detecting AdTracker related interactions~\cite{bhagavatula2014leveraging}. We have considered for each potential AdTracker sub-domain two types of content features to extract: "engineered features", and "keywords". The engineered features are four basic metrics like the URL lengths (that is generally large for AdTrackers), the number of times of repetition of special keys like "\&", "=" or "?," that are used to delimit values that are uploaded to AdTracker servers.  We also add the type of nodes {\em e.g.}, Js, Iframe, {\em etc} as an engineered feature.

Keywords are the second category of content features that are extracted from URLs.
We have first extracted all potential third-party AdTracker URLs and split them by special characters (e.g., "/", "?", "\&", "=", "." and "-") to obtain their keywords.
In order to evaluate the relevance of each candidate keyword, we used the well-known Term Frequency–Inverse Document Frequency (TF-IDF) statistics \cite{TF-IDF}. This metric is frequently used to evaluate how important a keyword is to a document in a collection. The TF-IDF score for a term $t$ in a document $d$ in a collection of documents  $\mathcal{D}$ is defined as :
\begin{equation*}
\textstyle{\mathrm{TF-IDF}(t,d,D) = \log({1+ f_{t,d})} \cdot  \log\left(\frac{|\mathcal{D}|}{1+|(d \in \mathcal{D}: t \in d)| }\right)},
\end{equation*}
where $f_{t,d}$ is the number of times that term $t$ occurs in the document $d$.  $|\mathcal{D}|$ is the number of documents and $|(d\in D:t\in d)|$ is the number of documents that contain the term $t$. We derived for each potential keyword its TF-IDF assuming that aggregation of URLs of a sub-domain is a separate document. We only saved the 1000 most valuable keywords and used them for building a content feature vector.
We show in Fig. \ref{fig-keyword} the average percentage of occurrence of the keywords with the largest TF-IDF in the AdTracker and the non AdTracker groups. It can be observed that some keywords are more prevalent in AdTrackers, but they are also present in a relatively large percentage of non-AdTrackers sub-domains, showing that content features are not enough to do the classification.

\subsection{\textbf{Structural Features}}
In addition to content features, we extracted, for each node in the {\em Wide-AdGraph}, some structural features using the Refex algorithm ~\cite{henderson2011s}. This is a recursive procedure to extract complex features from basic ones. We show in Fig. \ref{fig-refex} the flow of computing representation for a particular node. In the first step, the representation of each node is initialized with 7 basic features: 
\begin{itemize}
    \item (1-3): the in-degree, out-degree and overall degree of the node,
    \item (4): the "ego interconnectivity", {\em i.e.}, the number of edges interconnecting direct node neighbors, 
    \item (5): the "ego out-degree", {\em i.e.}, the number of edges connecting neighbors to other parts of the graph,
    \item (6-7): the direct and indirect coverage
\end{itemize}
In step 2 each node uses its neighbors' features, aggregates them to its own features, and appends them to initial features. This process continues until the features of each node contain information from the whole graph (calculated relative to itself). Furthermore, in our implementation, we used a pruning function that evaluates the correlation between features and drops those with high correlation.

\begin{figure}[t]
      \begin{minipage}{0.32\textwidth}
     \centering
     \includegraphics[width=\linewidth]{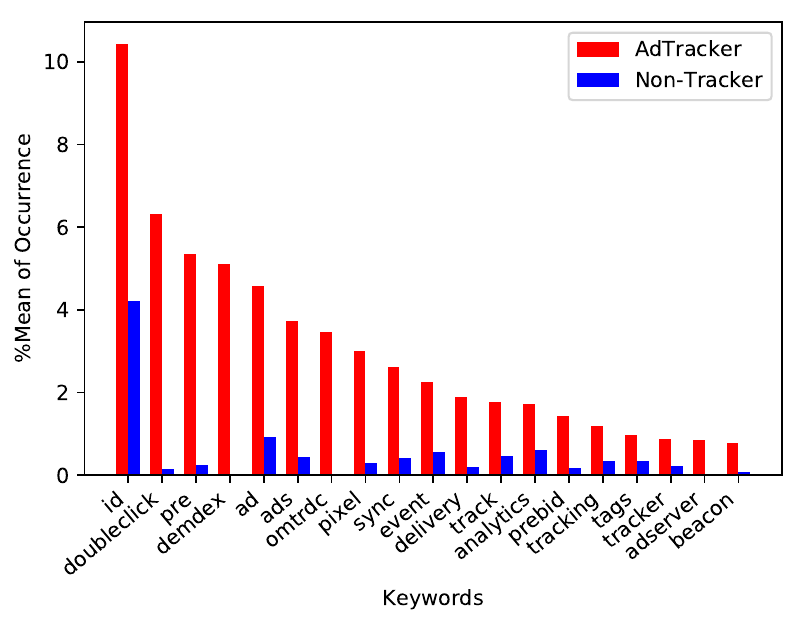}
     \caption{Average occurrence~(\%) of the keywords with the largest TF-IDF}\label{fig-keyword}
  \end{minipage}
  \begin{minipage}{0.34\textwidth}
     \centering
     \includegraphics[width=\linewidth]{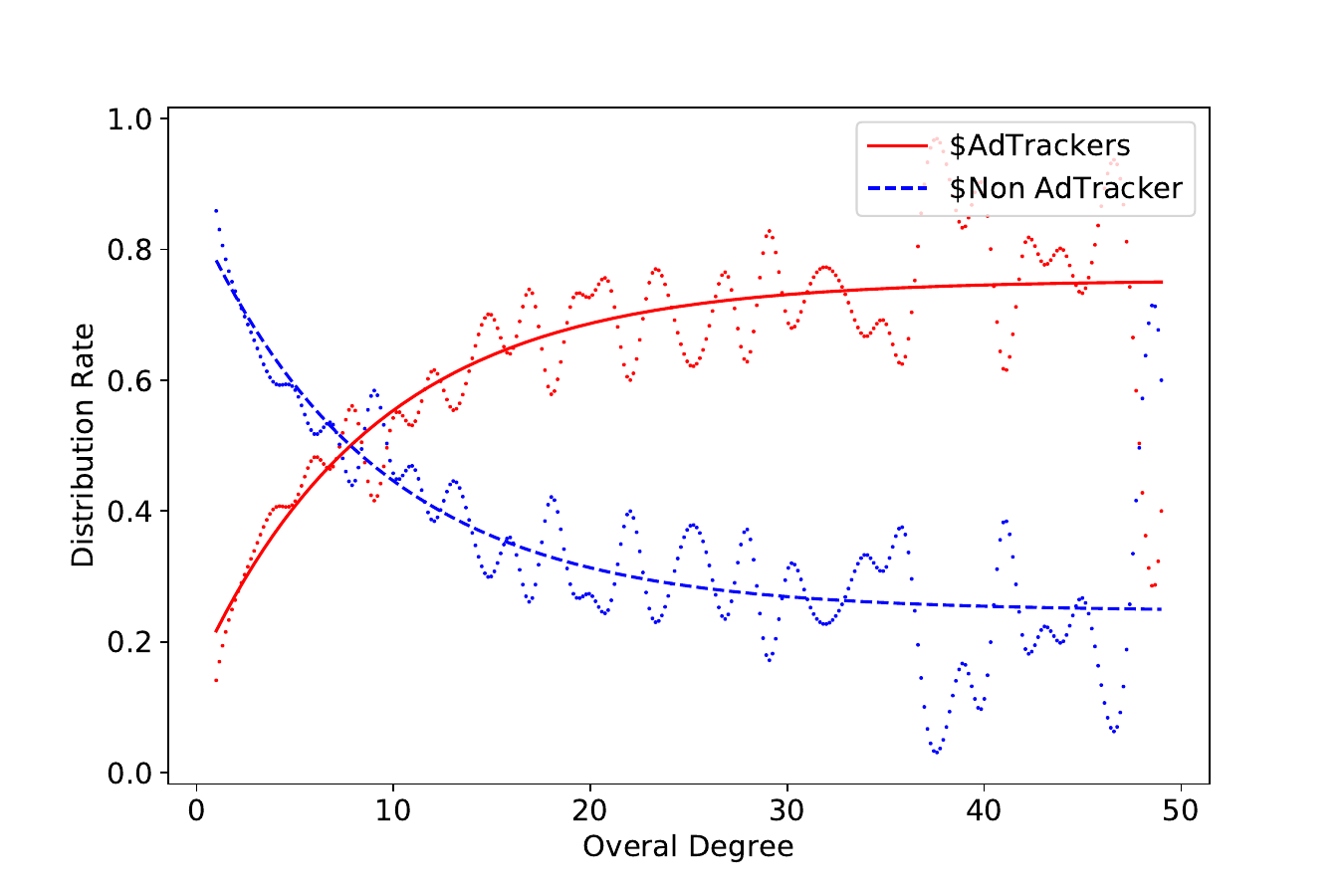}
     \caption{Distribution rate of overall degree}\label{fig-degree}
  \end{minipage}\hfill
  \begin{minipage}{0.34\textwidth}
     \centering
     \includegraphics[width=\linewidth]{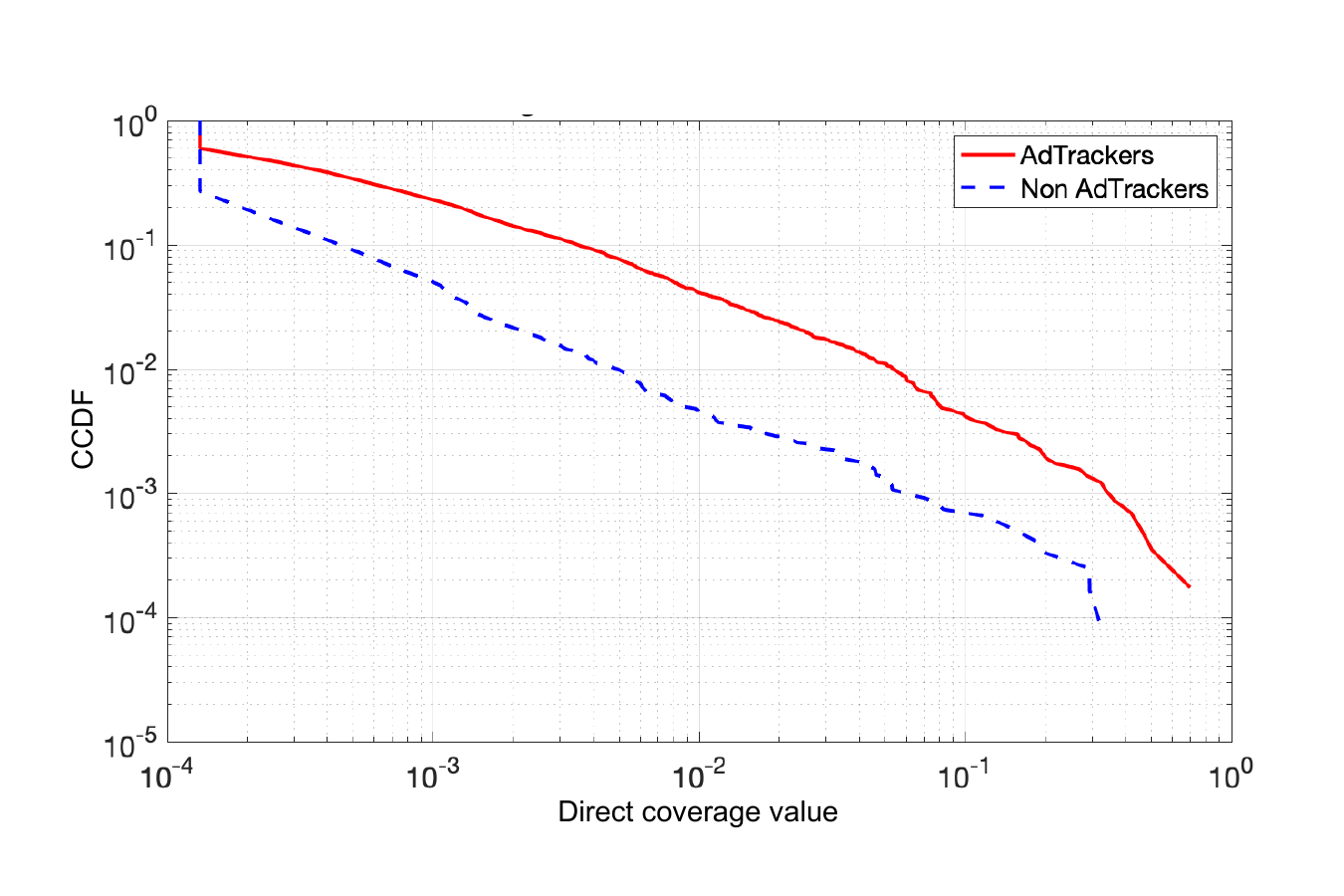}
     \caption{CCDF of direct coverage}\label{fig-ccdf}
  \end{minipage}
    \begin{minipage}{0.32\textwidth}
     \centering
     \includegraphics[width=\linewidth]{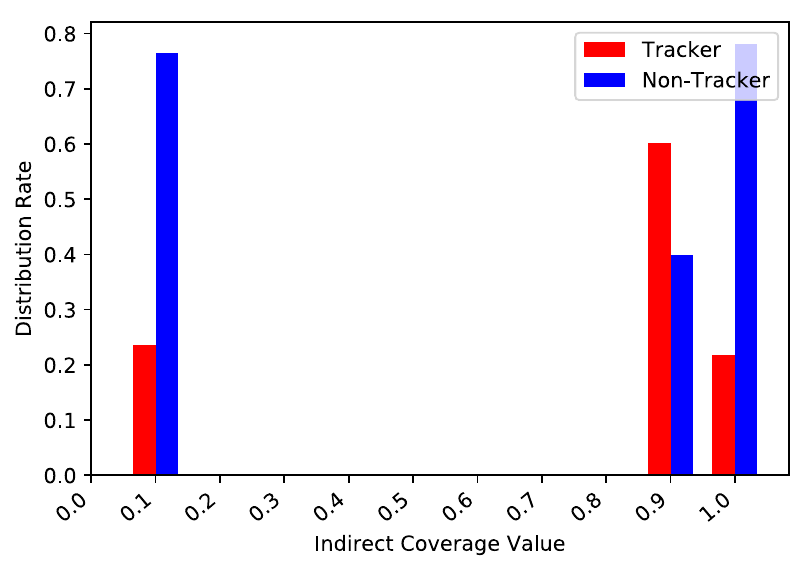}
     \caption{Distribution rate of indirect coverage}\label{fig-path}
  \end{minipage}
\end{figure}

We show in Fig. \ref{fig-degree}, Fig. \ref{fig-ccdf}, and Fig. \ref{fig-path} a preliminary analysis of 3 of the 7 features we are using as basic features. Fig. \ref{fig-degree} shows the proportion of third-party sub-domains that are labeled as AdTrackers and non-AdTrackers for each value of the degree. As it can be observed, sub-domains with a global degree less than 10 are mostly in the non-AdTracker group, while the proportion inverses for degrees larger than 10. However, the plot also shows that the overall degree is not discriminative enough as 20\% of very high degree nodes are still in the non-AdTracker category. We observe the same pattern for the in-degree and out-degree.

In Fig. \ref{fig-ccdf} the CCDF of the direct coverage obtained over the two classes of sub-domains. As can be seen, the direct coverage is more discriminative than the degree. For example, only 0.38\% of non-AdTrackers have direct coverage larger than 0.01, while 3\% of AdTrackers have such direct large coverage. Nonetheless, a relatively large proportion of AdTrackers still have direct coverage that is small. Fig. \ref{fig-path} makes the analysis for indirect coverage. A relatively large proportion of third-parities have indirect coverage between 0.8 to 0.9, while the proportion inverses for indirect coverage of less than 0.1 and larger than 0.9.

\begin{table*}[t]
\centering
\caption{Classification Result}
\label{tab-acc}
\begin{tabular}{|c|cc|cc|cc}
\hline
\multicolumn{1}{|l|}{} & \multicolumn{2}{c|}{\begin{tabular}[c]{@{}c@{}}(a) \\ Biased Performance\end{tabular}} & \multicolumn{2}{c|}{\begin{tabular}[c]{@{}c@{}}(b) \\ Unbiased  Performance\end{tabular}} & \multicolumn{2}{c|}{\begin{tabular}[c]{@{}c@{}} (c) \\ Corrected Unbiased Performance\end{tabular}} \\ \hline
                       & Precision                                   & Recall                                  & Precision                                    & Recall                                   & Precision                        & \multicolumn{1}{c|}{Recall}                           \\ \hline
AdTrackers             & 99.2\%                                      & 95.5\%                                  & 90.2\%                                       & 87.7\%                                   & \textbf{96.6\%}                  & \multicolumn{1}{c|}{\textbf{88.4\%}}                  \\ \hline
Non-AdTracker         & 88.9\%                                      & 97.8\%                                  & 83.3\%                                       & 86.6\%                                   & 83.3\%                           & \multicolumn{1}{c|}{\textbf{94.9\%}}                  \\ \hline
Macro Avg              & 94.0\%                                      & 96.7\%                                  & 86.8\%                                       & 87.1\%                                   & \textbf{90.0\%}                  & \multicolumn{1}{c|}{\textbf{91.7\%}}                  \\ \hline
Accuracy               & \multicolumn{2}{c|}{96.1\%}                                                           & \multicolumn{2}{c|}{87.2\%}                                                             & \multicolumn{2}{c|}{\textbf{90.9\%}}                                                     \\ \hline
\end{tabular}
\end{table*}

\section{Classifying AdTrackers}\label{sec-classification}
The aim of our approach is to develop a classifier that given "structural" and "content" features extracted from a given sub-domain (or a URL) predicts whether that sub-domain is an AdTracker or a benign redirection. In this section, we will describe the supervised learning algorithm that will generate a classifier. 
\subsection{Model Selection}
We tested two methods for building this classifier: a Deep Neural Network, and a random forest model. As the results of the random forest are slightly better and it is less computationally expensive, we report here only the results of the random forest model. A random forest is an ensemble learning method for classification that operates by constructing a multitude of decision trees at training time and results the class that is the mean prediction of the individual trees. Random decision forests correct for decision trees' habit of overfitting to their training set. For this reason, random forests are among popular first approaches for the classification of complex multi-dimensional data that cannot be fitted by simple regression.

\subsection{Model Fitting}
The learning set used for model fitting is generated by using third-party sub-domain nodes in the {\em Wide-AdGraph}. We removed from the learning set all sub-domains with in-degree less than 2, as an object that only used by only one website might be considered to be part of that even if their domains are different. For instance, \textit{geforce.com} calls \textit{nvidia.com/*/geforce} and many of the third-party of \textit{geforce.com} and \textit{nvidia.com} work the same. This filtering step removed 14394 sub-domains out of the initial 33373 potentially third-party sub-domains, and  18979 sub-domains are retained. We selected 80\% of them as the learning set and put aside the remaining 20\% for evaluation. 

We labeled the sub-domains in the learning set with labels coming from most popular filter list\footnote{We have used the \textit{adblockparser} tool to label URLs~\cite{adblockpaerser}}: Easy List~\cite{easylist}, Easy Privacy~\cite{easyprivacy}, Squid Blacklist~\cite{squid}, Fanboy's Social Blocking List~\cite{fanboy-social} and Fanboy's Annoyance List~\cite{fanboy-annoyance}. These filter lists include more than 100K rules to detect URLs attached to an AdTracker. We label a sub-domain as  "AdTrackers" if at least one of the URLs belonging to this sub-domain has been blocked by the filter lists.

The random forest model is therefore fitted over a vector of 1061 attributes for each potential third-party sub-domain: 1000 keywords resulting from content features, 5 engineered features, and 56 structural features extracted from the {\em Wide-AdGraph}. These feature vectors are generated by combining the features of all interaction URLs within that sub-domain (document). We use hyperparameter tuning with 10-fold cross-validation to choose the number of trees. The training set contained 58\% of entries labeled as "AdTracker" and 42\% of "non-AdTracker" ones. The fitting needed around 10 sec to be finished. We present the results of the fitted random forest tree in the next section.

\section{Evaluation}\label{sec-evaluation} 
In this section, we report and analyze the results of our trained classifier. The fitted random forest was applied to our evaluation set containing 3796 sub-domains. We present the results in Table \ref{tab-acc}. Before going further we clarify the definition of false positives and false negatives. 

\begin{itemize}
 \item \textbf{False positives}: the random forest classifier predicts that a particular sub-domain is an AdTrackers, while the filter lists have labeled it as not AdTrackers.
  \item \textbf{False negatives}: the random forest classifier predicts that a particular sub-domain is a non-AdTrackers, while the filter lists labeled it as AdTrackers.
\end{itemize}
With these definitions, we can calculate for each class of sub-domains a precision and a recall value resulting in a confusion matrix. We can also calculate the performance measures for the classifier.

Another major point is relative to the way of calculating the performance metrics. The popularity of websites is highly skewed, {\em i.e.}, in the normal operation of a browser some target domains will appear much more frequently than others. Moreover, some AdTrackers are much more prevalent than others, {\em e.g.}, Google Analytic is appearing much more frequently than some less used AdTrackers. These two nonuniformity introduce a bias when we calculate the performance over the flow of incoming websites, {\em i.e.}, the impact on the performance of detecting correctly a popular AdTracker is much higher than detecting a less popular one. A less biased approach will consist of giving the same weight to all sub-domains, whether they are popular or not, and to evaluate the performance over all potential AdTrackers rather than doing it over the set of AdTrackers observed during a browsing experience. Unfortunately, almost all papers in the literature adopt the first approach, rather than the less biased second approach. This results into difficulty in interpreting and comparing the high-level of performance reported in the literature, as we do not know if the very good performance is resulting from the popularity bias, or from the good performance of the classifier. Moreover, the second approach tends to reduce classification performance. This bias is also extended to techniques that use individual graphs for each website instead of a single global graph like in \cite{sjosten2020filter,iqbal2020adgraph}. In such cases, the same tracker appears several times for different websites, its metrics repeat several times in the learning set, and biases strongly the learning algorithms.

In this paper, we decided to report both performance metrics: the biased and the unbiased ones. Table \ref{tab-acc}-a shows the biased performance  indicating that we have been able to obtain 99.2\% precision on AdTrackers observed during our crawling experience on the Alexa 10K top sites, and that we have achieved an average accuracy
of  96.1\%. These values are very much aligned with other performance reported in the literature, in state-of-the-art papers~\cite{sjosten2020filter,cozza2020hybrid,iqbal2020adgraph}. However, we report also in Table \ref{tab-acc}-b, the unbiased performance we obtain. As expected this performance underperforms the biased one. Nevertheless, we achieve 87.2\% of accuracy over all sub-domains.

We did a manual review of false-positive cases, where the random forest predicted an AdTracker and the filter lists did not. Our review methodology includes looking at the content of the Images and Videos, searching URLs on the internet and checking the related company websites and their services, looking at their partners, checking if they have cookie transmission or fingerprint functionalities. Among these 220 false-positive cases we found 138 sub-domains (62\% of false positives) that are in fact AdTrackers that have not been detected by filter lists. Meaning that the random forest model detected some new AdTrackers compared to the filter lists. This validates the usability of our methodology to update automatically filter lists. After correcting the labels of these AdTrackers, table~\ref{tab-acc}-c lists the  recalculated performance. The corrected performance is now 90.9\% of accuracy and 96.6\% precision for AdTrackers. We did also an analysis of the false negatives sub-domains and observed that they were all with low popularity and have been referred by a limited number of other AdTrackers. Our works can be extended to handle some of these sub-domains by adding new features from JavaScript APIs, and perceptual-based features ~\cite{sjosten2020filter,storey2017future} as well.

\section{Related Works}\label{sec-related}
Many researchers are currently engaged in designing ad and trackers detectors using machine learning techniques. In the following, we review these research works. 

\subsection{HTTP-Based Approaches}
These approaches try to find HTTP-based patterns to classify ads and trackers. For example, Bhagavatula et al.~\cite{bhagavatula2014leveraging} used characteristics of the structure of the URL to train a machine learning model that classifies ad-related URLs from non-ad-related URLs. Gugelmann et al.~\cite{gugelmann2015automated} used HTTP traffic traces attributes like the size and the number of HTTP requests to train a machine learning model that classifies ads and analytic services. They evaluated their model on real traffic traces of a campus network and found new privacy-intrusive services. Yu et al.~\cite{yu2016tracking} also analyzed HTTP requests and used a method to find tracker domains by looking for the third-party objects that receive similar unique patterns from a large number of first-parties. 

\subsection{JavaScript Approaches}
In \cite{ikram2017towards}, Ikram et al. used supervised machine learning by using syntactic and semantic features from JavaScript files to predict that a JavaScript is malicious or not.
Wu et al. \cite{wu2016machine} used machine learning to classify whether a JavaScript code unit is tracking related or not. Also, they identified HTTP requests generated by the third-party tracking JavaScripts. In \cite{kaizer2016towards}, Kaizer and Gupta trained a classifier with features related to JavaScript access, cookie access, and URL information to detect machine-based trackers. While these approaches achieve high-grade accuracy, most of them are not robust against obfuscation techniques \cite{xu2012power-obfuscation}, \cite{xu2013jstill-obfuscation}, \cite{le2017towards-obfuscation}.

\subsection{Extended Approaches}
In order to increase accuracy and robustness, researchers used multiple types of objects. For example, Cozza et al. proposed a hybrid method that used both machine learning and filter-lists depending on whether an AdTracker is new or not~\cite{cozza2020hybrid}. They applied their model to JavaScript and HTTP requests. Authors in \cite{bau2013promising}
used HTML and HTTP information to build a graph of resource-hosting domains from the Document Object Model (DOM) hierarchy of each website. 
They leverage machine learning to classify web tracking domains.
In \cite{iqbal2020adgraph}, Iqbal et al. proposed \emph{Adgraph} and designed a machine learning classifier that takes as input the structural and content features for each website \emph{individually}. 
Authors in \cite{sjosten2020filter} extended the Adgraph work by adding a perceptual classifier to generate rules to update filter lists for under-served regions (non-English regions). This is similar to our approach as we also use an offline process to find more hidden trackers from a larger perspective to update the filter lists.
But as we discussed in section~\ref{subsec-adgraph} our method differs in two fundamental aspects. The first one is that our model has no \emph{bias} and the second one is that our algorithm builds a \emph{single} holistic graph, not separate graphs.

To the best of our knowledge, our paper is the first try to comprehensively collect and analyze the interactions between the objects on web pages in large scale to detect ads and trackers. As our evaluations have shown, we are able to find ads and trackers that we had never seen before in the training set.

\section{Conclusion}\label{sec-conclusion}
Third-party AdTracker services are very dynamic and may change their behavior repeatedly. As a result, many AdTracker detection mechanisms become vulnerable, especially the ones working only based on URLs patterns or code analysis. In this paper, we proposed a holistic approach that considers conventional URL features, alongside the behavior of the third-parties in the AdTracker ecosystem.
Unlike the most state-of-the-art machine learning approaches that use separate dependency graphs for individual websites, our proposed method builds a large scale {\em Wide-AdGraph} by properly merging the dependency graphs of individual websites.
Using the {\em Wide-AdGraph}, we can encode the third-parties' specific roles in the whole network into structural (graph-based) features. Using strong content and structural features, we have trained a random forest model to classify third-party nodes into AdTracker and non-AdTracker categories. The resulted classifier has shown a high degree of robustness to the dynamic behavior of third-party nodes,
reached to 96.1\% biased and 90.9\% unbiased accuracy, and uncovered hundreds of new AdTrackers that are not hit by filter-lists at the time of this writing. 

\balance
\bibliographystyle{ACM-Reference-Format}
\bibliography{wideadgraph}

\end{document}